# Atomically thin quantum light emitting diodes

Carmen Palacios-Berraquero[†1], Matteo Barbone[†2], Dhiren M. Kara[1], Xiaolong Chen[2], Ilya Goykhman[2], Duhee Yoon[2], Anna K. Ott[2], Jan Beitner[1], Kenji Watanabe[3], Takashi Taniguchi[3], Andrea C. Ferrari[2*] and Mete Atatüre[1*]

[1]*Cavendish Laboratory, University of Cambridge, JJ Thomson Ave., Cambridge CB3 0HE, UK*

[2]*Cambridge Graphene Centre, University of Cambridge, Cambridge CB3 0FA, UK*

[3]*Advanced Materials Laboratory, National Institute for Materials Science, Tsukuba, Ibaraki 305-0034, Japan*

[†]These authors contributed equally to this work.

[*]Corresponding authors: A.C.F. (acf26@cam.ac.uk) and M.A. (ma424@cam.ac.uk).

## Abstract

Transition metal dichalcogenides are optically active layered materials providing potential for fast optoelectronics and on-chip photonics. We demonstrate electrically driven single-photon emission from localised sites in tungsten diselenide and tungsten disulphide. To achieve this, we fabricate a light emitting diode structure comprising single layer graphene, thin hexagonal boron nitride and transition metal dichalcogenide mono- and bi-layers. Photon correlation measurements are used to confirm the single-photon nature of the spectrally sharp emission. These results present the transition metal dichalcogenide family as a platform for hybrid, broadband, atomically precise quantum photonics devices.

**Introduction**

Incorporating single-photon sources into optoelectronic circuits is a key challenge to develop scalable quantum-photonic technologies. Despite a plethora of single-photon sources reported to-date, all-electrical operation, desired for systems integration, is reported for only a few[1–4]. Layered materials (LMs) offer novel opportunities for next-generation photonic and optoelectronic technologies[5,6], such as lasers[7,8], modulators[9,10] and photodetectors[11], and are compatible with the silicon platform[12].

The attractiveness of single-photon sources in LMs[13–18] stems from their ability to operate at the fundamental limit of few-atom thickness, providing the potential to integrate into conventional and scalable high-speed optoelectronic systems[19,20]. Transition metal dichalcogenides (TMDs), being optically active layered semiconductors, are particularly suitable for developing quantum-light generating devices.

Here, we demonstrate that LMs enable all-electrical single-photon generation over a broad spectrum. We use a light emitting diode (LED) realised by vertical stacking of LMs and achieve charge injection into the active layer containing quantum emitters. We show that quantum emitters in tungsten diselenide ($WSe_2$)[13,14,16,15,17] can operate electrically. We further report all-electrical single-photon generation in the visible spectrum from a new class of quantum emitters in tungsten disulphide ($WS_2$). Our results highlight the promise of LMs as a new platform for broadband hybrid all-integrated quantum-photonic circuits.

**Results**

**Design and operation of a vertical TMD-based LED.** We realise an LED based on a single tunnelling junction made of vertically stacked LMs (see Methods and Supplementary Note 1). Figure 1a is a typical optical microscope image of such a device. From bottom to top, three layers form a heterostructure on a silicon/silicon dioxide (Si/SiO$_2$) substrate: a single layer of graphene (SLG), a thin (2-6 atomic layers) sheet of hexagonal boron nitride (hBN), and finally a mono- or bilayer of TMD, such as WSe$_2$. The WSe$_2$, exfoliated from a naturally p-doped bulk crystal, is the optically active layer hosting single-photon sources. Metal electrodes provide electrical contact to the SLG and TMD layers. To obtain electroluminescence (EL), we inject electrons from the SLG to the p-doped WSe$_2$ through the hBN tunnel barrier (see Supplementary Fig. 9 for current-voltage characteristics of the devices). A vertically stacked heterojunction allows for EL from the whole device area, unlike lateral Schottky junction or split-gate p-n junction devices[21–23], and provides the benefit of atomically precise interfaces and barrier thicknesses[24]. We leave the optically active TMD layer exposed at the top of the device purposefully to offer interfacing with other systems.

Figure 1b illustrates the operational concept of our LED. At zero bias between the SLG and the monolayer TMD contacts, the Fermi energy ($E_F$) of the system is constant across the heterojunction, preventing net charge flow (current) between the layers (Fig. 1b, top). A negative bias applied to the SLG electrode raises the SLG $E_F$ above the minimum of the conduction band ($E_C$) of grounded WSe$_2$ and thus electrons tunnel from the SLG into the monolayer WSe$_2$. This initiates photoemission through radiative recombination between the tunnelled electrons and the holes residing in the optically active WSe$_2$ area (Fig. 1b, bottom). Figure 1c compares the EL and photoluminescence (PL) spectra from this monolayer-WSe$_2$-based LED device for two operation temperatures, room temperature (RT) and 10 K (see Methods and Supplementary Fig. 10 for measurement setup). PL at RT is given by the black curve in the lower panel with a broad peak at 750 nm corresponding to the monolayer WSe$_2$ unbound neutral exciton emission, X$^0$ [25]. Under electrical excitation the main peak is shifted ~20 nm to longer wavelengths, which is commensurate with the charged exciton, $X^-$ [26], as shown in the blue curve. This charge state attribution was confirmed with a separate back-gated device, which allowed tuning of the WSe$_2$ Fermi level. The black and blue spectra in the upper panel of Fig. 1c show the device's PL and EL emission at 10 K, respectively. Due to the increased bandgap at low temperatures, the unbound exciton emission is shifted to shorter wavelengths by ~30 nm[27]. Consistent with recent reports[13–17,27], extra structure arising from localised exciton state emission (L) appears at longer wavelengths in the PL spectrum. Critically, these features are also visible under electrical excitation. In the low current regime (<1 µA for this device) they dominate the EL spectrum, as shown in Fig. 1c, indicating that localised exciton states respond more efficiently to charge injection than the delocalised ones.

**Electrically-driven quantum light in a WSe$_2$-based LED.** Figure 2a is a spatial map of integrated EL from a WSe$_2$-based LED device at 10K. The active region of this device consists of adjacent monolayer and bilayer active areas, both in contact with the ground electrode. The brighter area in Fig. 2a corresponds to the bilayer, suggesting that most of the injected current flows through this region (see Supplementary Fig. 11). In addition to the spatially uniform light emission from delocalised excitons, we observe quantum LED (QLED) operation in the form of highly localised light emission from both the monolayer and the bilayer WSe$_2$, identified by the dotted circles (Fig. 2a). These localised states lie within the bandgap of WSe$_2$ and therefore emit at lower energies (longer wavelength) with respect to the bulk exciton emission (see Fig. 2b)[13–17]. Figure 2c shows example emission spectra from these sites, where the top (bottom) spectrum belongs to a quantum emitter in the monolayer (bilayer) WSe$_2$ section. We observe spectrally isolated peaks from multiple locations in most devices with linewidths ranging between 0.8 nm and 3 nm. We see on average 1-2 emitters per ~40 µm$^2$ active device area. Electrically excited narrow lines coming from bilayer WSe$_2$ regions are typically redshifted with respect to those coming from the monolayer regions[28]. The emission peaks of Fig. 2c are unpolarised, and the fine structure splitting reported in PL experiments (~0.3 nm[13–17]) is not resolvable due to the broader linewidths we observe in EL. On the timescale of seconds, most emitters show spectral wandering, of up to 2 nm, similar to that seen in our PL measurements. Gating and encapsulation of the active layer should aid in reducing the broad linewidths observed here, which we attribute to charge noise in the device. Slow spectral fluctuations can further be reduced through active feedback, for example via the direct current Stark shift[29,30]. A fraction of the quantum emitters display blinking, discrete spectral jumps, or multiple spectral lines at similar timescales (see Supplementary Fig. 12). We also see that, as in PL, the electrically driven emitters display robust operation, withstanding multiple (3 to 5) cooling/heating cycles and several hours of measurement under uninterrupted current flow.

Figure 2d plots the current dependence of the integrated EL intensity from a quantum emitter, as well as from the unbound monolayer WSe$_2$ excitons. The latter shows a predominantly linear relationship between emission intensity and injected current; however, EL emission from the quantum emitter shows clear saturation as a function of current, a universal behaviour seen with single-photon sources[31] (see Supplementary Fig. 13 for a plot of the unbound exciton and quantum emitter spectra as a function of current). Figure 2e shows the measured intensity-correlation function, $g^{(2)}(\tau)$, of the integrated-EL emission from a WSe$_2$-based QLED using a standard Hanbury Brown and Twiss interferometer (see Methods). The value of the normalised $g^{(2)}(0)$ drops to 0.29±0.08, well below the threshold value of 0.5 for a single-photon source[1]. We note that these data are not corrected for background emission within the broad spectral window of detection or for the photon-counting detector dark counts, which together contribute to the non-zero value of $g^{(2)}(0)$. While our results manifest the single-photon nature of the electrically generated emission as proof-of-concept, the immediate usability of these quantum emitters as single-photon sources would benefit from spectral filtering to supress the background emission. Further technical improvements, such as optimised designs for charge injection, may be possible once the nature of these emitters is identified.

**All-electrical generation of single-photon emission in $WS_2$.** In TMD-based quantum emitters the host material influences the quantised energy levels and consequently their emission wavelength. Therefore, in order to obtain single-photon emission at a complementary part of the spectrum, we replace the monolayer of $WSe_2$ with $WS_2$ (exfoliated from an n-doped bulk crystal) as the active layer; the rest of the QLED device structure remains unchanged. Figures 3a and 3b display the spatial maps of integrated EL emission from a $WS_2$-based QLED device at high- (0.665 µA) and low- (0.570 µA) current injection, respectively. At high current, the emission intensity is spatially uniform in the monolayer. At low currents however, a spatially localised emission site dominates, indicating that $WS_2$ also hosts localised quantum emitters. Figure 3c shows the EL spectrum as a function of injection current, demonstrating that the low current (~0.570 µA) regime leads to a narrow (~4 nm) emission at 640 nm, a line cut (in blue) of which is in the bottom right panel. Fig. 3c (upper right plot) also shows how the EL spectrum is broadened significantly when driven strongly at an injection current of 1.8 µA. The EL at 640 nm lies within the spectral region of an emission band that appears, in addition to the unbound exciton emission, at low temperature (<10 K) under optical excitation (see Supplementary Figs 14 and 15 for details). Operating in the low current range ensures that the full EL spectrum is dominated by single-photon emission from the quantum emitter, obviating any need for tailored spectral filtering. The intensity-correlation measurement for EL in this regime, without spectral filtering, yields the $g^{(2)}(\tau)$ data shown in Fig. 3d. Similar to the WSe$_2$ emitters, the uncorrected, but normalised, $g^{(2)}(0)$ falls below 0.5 to 0.31±0.05 indicating that WS$_2$ supports stable QLED operation, generating single photons in the visible spectral range.

**Discussion**

Our TMD-based QLEDs rely on a single tunnelling heterojunction design, where a wide range of TMDs can be active materials. Other designs, employing a back gate to tune $E_F$ of the active TMD layer and providing electrostatic tuning of the EL emission spectrum, can enhance the versatility of these devices. One possibility is the deterministic control over the charging states of confined excitons[32], en route to spin control[33] and entangled photon generation[34]. We also note that the emission wavelength range for WSe$_2$ emitters can match rubidium transitions (~780 nm) for exploring quantum storage possibilities. Similarly, silicon-vacancy centres (~737 nm) and nitrogen-vacancy centres (~637 nm) in diamond can have matching transitions with the WSe$_2$ and WS$_2$ QLEDs, respectively, for interfacing hybrid quantum systems via distributed or on-chip photonic channels. Other TMDs are likely to yield similar results decorating different spectral windows. Our results offer promise to pursue these opportunities and demonstrate that layered materials are a platform for fully integrable and atomically precise devices for quantum photonics technologies.

**Methods**

**Device Fabrication.** We exfoliate the LMs on oxidised Si wafers by micromechanical cleavage of bulk crystals of graphite (from NGS Naturgrafit), TMDs (from HQgraphene) and hBN (hBN single crystals were grown by the temperature-gradient method under high pressure and high temperature)[35]. Mono-, bi- and few-layer samples are identified by a combination of optical contrast[36], Raman spectroscopy[37], PL, and atomic force microscopy (AFM) (see Supplementary Note 1). Single crystals are assembled into heterostructures via a dry-transfer technique[38]. A transparent stack comprising a glass slide, a polydimethylsiloxane (PDMS) layer attached to the glass and polycarbonate (PC) as an external film is mounted on a micromanipulator positioned under an optical microscope with a temperature-controlled stage. After adjusting the alignment and bringing the transfer stack into contact with the exfoliated TMD flakes, these are picked up due to their higher adhesion to PC. The process is repeated for the hBN tunnel barrier. Finally, after aligning and bringing in contact hBN and TMD on PC with exfoliated SLG on $Si/SiO_2$, the temperature is raised to ~100 °C, releasing the PC with hBN/TMD onto SLG. Then, the sample is soaked in Chloroform to dissolve the PC film, leaving the SLG/hBN/TMD heterostructure on the $Si/SiO_2$ substrate. Finally, Cr/Au (3/50 nm) contacts both to SLG and TMD are patterned by e-beam lithography following a standard lift-off process. Heterostructures are characterised by Raman spectroscopy to ensure no degradation (see Supplementary Notes 1-3 for details).

**Confocal microscope.** PL and EL measurements are performed using a home-built confocal microscope mounted on a three-axis stage (Physik Instrumente M-405DG) with a 5-cm travel range and 200-nm resolution for coarse alignment and a piezo scanning mirror (Physik Instrumente S-334) for high resolution raster scans (see Supplementary Fig.10 for a diagram of the optical setup). PL and EL are collected using a 1.7-mm working distance objective with a numerical aperture of 0.7 (Nikon S Plan Fluor 60x) and detected on a fibre-coupled single-photon-counting module (PerkinElmer SPCM-AQRH). A variable-temperature helium flow cryostat (Oxford Instruments Microstat HiRes2) is used to perform PL and EL measurements. A controlled bias is applied to the QLED devices by a source measurement unit (Keithley 2400) for

EL experiments. Intensity correlations from the Hanbury Brown and Twiss interferometer are recorded with a time-to-digital converter (quTAU). A double grating spectrometer (Princeton Instruments) is used for acquiring spectra. For PL measurements, the excitation laser (700 nm / 520 nm, Thorlabs MCLS1) is suppressed with a long pass filter (715 nm, Semrock FF01-715 / 550 nm Thorlabs FEL0550).

**Data Availability**

The data that support the findings of this study are available from the corresponding author upon request.

**Acknowledgements**

We acknowledge financial support from the EU Graphene Flagship (no. 604391), ERC Grants Hetero2D and PHOENICS, EPSRC Grants EP/K01711X/1, EP/K017144/1, EP/N010345/1, EP/M507799/1, EP/L016087/1, EP/M013243/1 and the EPSRC Cambridge NanoDTC, EP/G037221/1. We are grateful to J. Barnes, C. Le Gall and H. S. Knowles for technical assistance.

**Author contributions**

M.A., A.C.F. and I.G. devised and managed the project. K.W. and T.T. provided high-quality hBN material, M.B., X.C. and I.G. fabricated the devices, C.P.B., and D.M.K. performed the optical measurements, assisted by M.B., X.C. and I.G., and analysed the data. M.B., C.P.B., D.Y. and A.K.O. performed Raman, PL, optical contrasts and AFM measurements and analysis. J.B.

developed part of the data acquisition software for the confocal microscope. All authors participated in the writing of the manuscript.

Correspondence and requests for materials should be addressed to A.C.F. (acf26@cam.ac.uk) and M.A. (ma424@cam.ac.uk).

**Competing financial interests**

The authors declare no competing financial interests.

**Figure 1 | Design and operation of a TMD-based LED. a,** Optical microscope image of a typical device used in our experiments. The dotted lines highlight the footprint of the SLG, hBN and the TMD layers individually. The Cr/Au electrodes contact the SLG and TMD layers to provide an electrical bias. **b,** Heterostructure band diagram. The top illustration shows the case for zero applied bias and the bottom illustration shows the case for a finite negative bias applied to the SLG. Tuning the SLG Fermi level ($E_F$) across the TMD conduction band edge ($E_C$) allows electron tunnelling from the SLG to the TMD, resulting in light emission via radiative recombination of the electrons with the holes residing in the p-doped TMD layer. The appearance of valence-band holes below the Fermi level is due to the natural p-doping of $WSe_2$. **c,** An example of layered LED emission spectra for an optically active layer of $WSe_2$. Top (bottom) spectra correspond to 10 K (RT) operation temperature, where the black and blue spectra are obtained by optical excitation and electrical excitation, respectively.

**Figure 2 | WSe$_2$-based QLED operation in the near infrared spectrum. a,** A raster-scan map of integrated EL intensity from monolayer and bilayer WSe$_2$ areas of the QLED for an injection current of 3 µA (12.4 V). The dotted circles highlight the sub-micron localised emission in this device. **b,** A schematic energy band diagram, similar to that in Fig. 1b, including the confined electronic states of the quantum dots. EL emission from quantum dots typically starts at lower bias than the conventional LED operation threshold. **c,** Typical EL emission spectra for quantum dots in monolayer (top) and bilayer (bottom) WSe$_2$. The shaded area highlights the spectral window for LED emission due to bulk WSe$_2$ excitons, while QLED operation produces spectra at longer wavelengths. **d,** Comparison of the integrated EL intensity for the WSe$_2$ layer and for a quantum dot as a function of the applied current. The linear increase in WSe$_2$ layer EL contrasts with the saturation behaviour of the QLED emission. **e,** Intensity-correlation function, $g^{(2)}(\tau)$, for the same emitter displaying the antibunched nature of the EL signal, $g^{(2)}(0) = 0.29\pm0.08$, and a rise-time of 9.4±2.8 ns.

**Figure 3 | WS$_2$-based QLED operation in the visible spectrum.** A raster-scan map of integrated EL intensity from the monolayer WS$_2$ area of the device: **a,** at 0.665 µA injection current (bias 2.08 V), where the emission is delocalised and roughly uniform, and **b,** at 0.570 µA (1.97 V), where the highly localised QLED emission dominates over the unbound WS$_2$ exciton emission. **c,** A map of the EL spectrum, displaying the evolution of the WS$_2$ QD emission spectrum as a function of current. The spectrum at the top (bottom) of the panel is a line cut for injection current of 1.8 µA (0.578 µA). **d,** Intensity-correlation function, $g^{(2)}(\tau)$, for the same quantum dot displaying the antibunched nature of the EL signal, $g^{(2)}(0) = 0.31\pm0.05$ and a rise-time of 1.4±0.15 ns.

# Figure 1

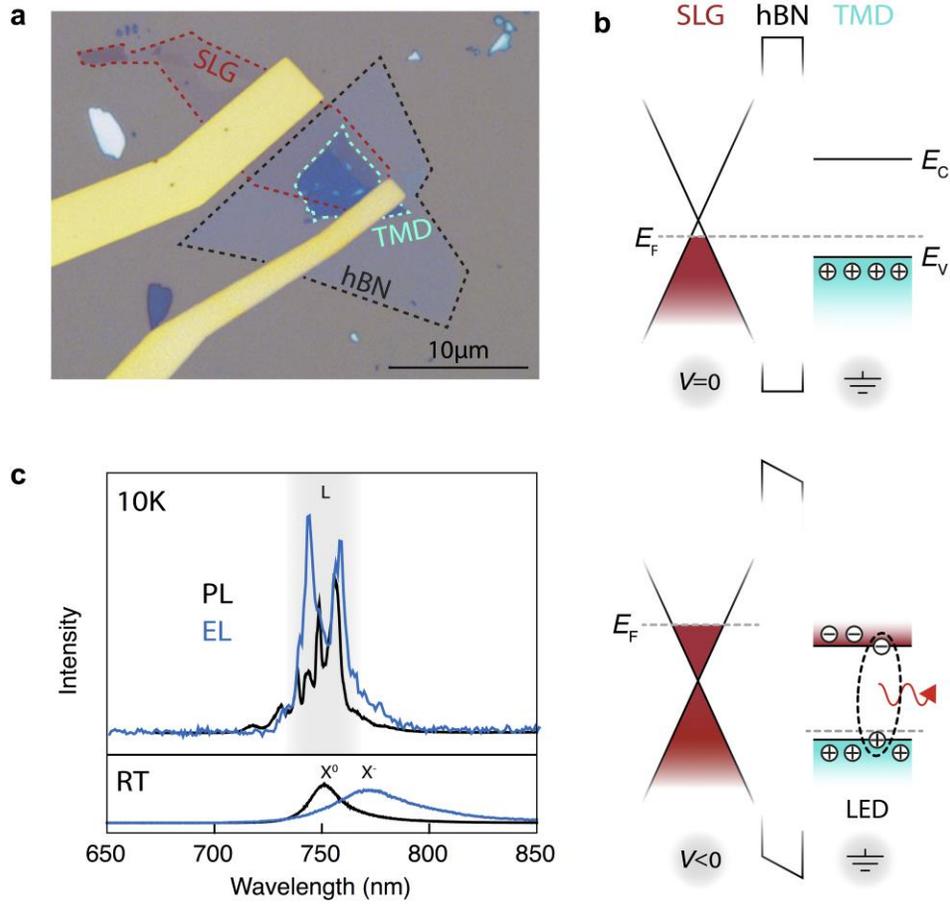

# Figure 2

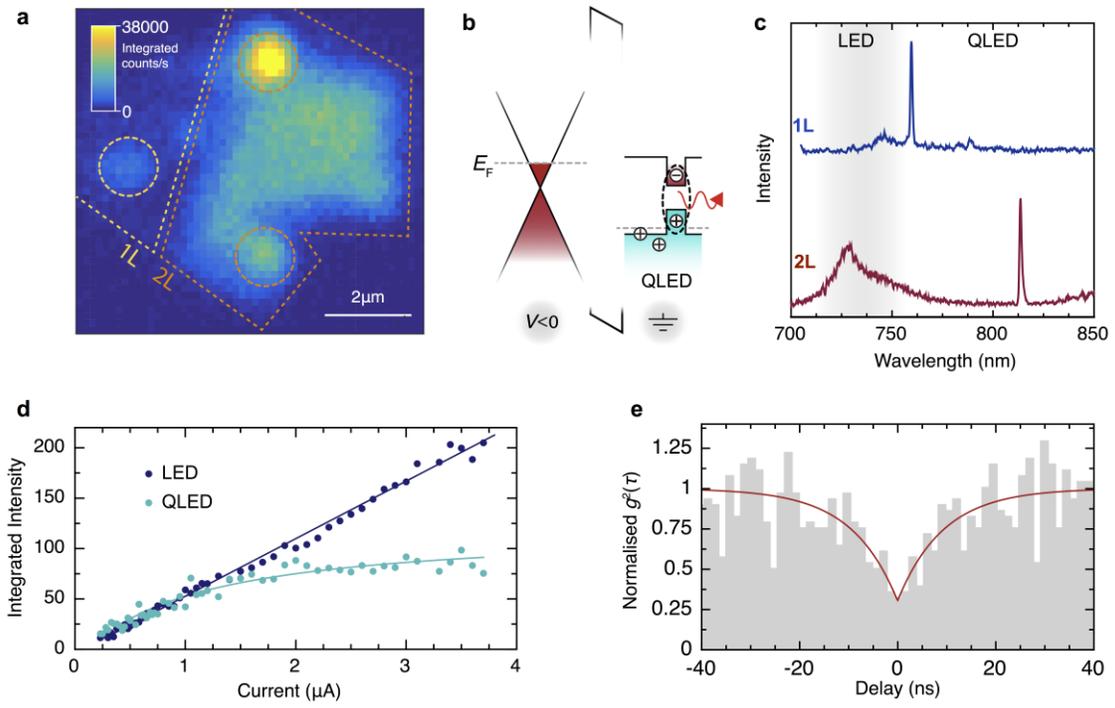

**Figure 3**

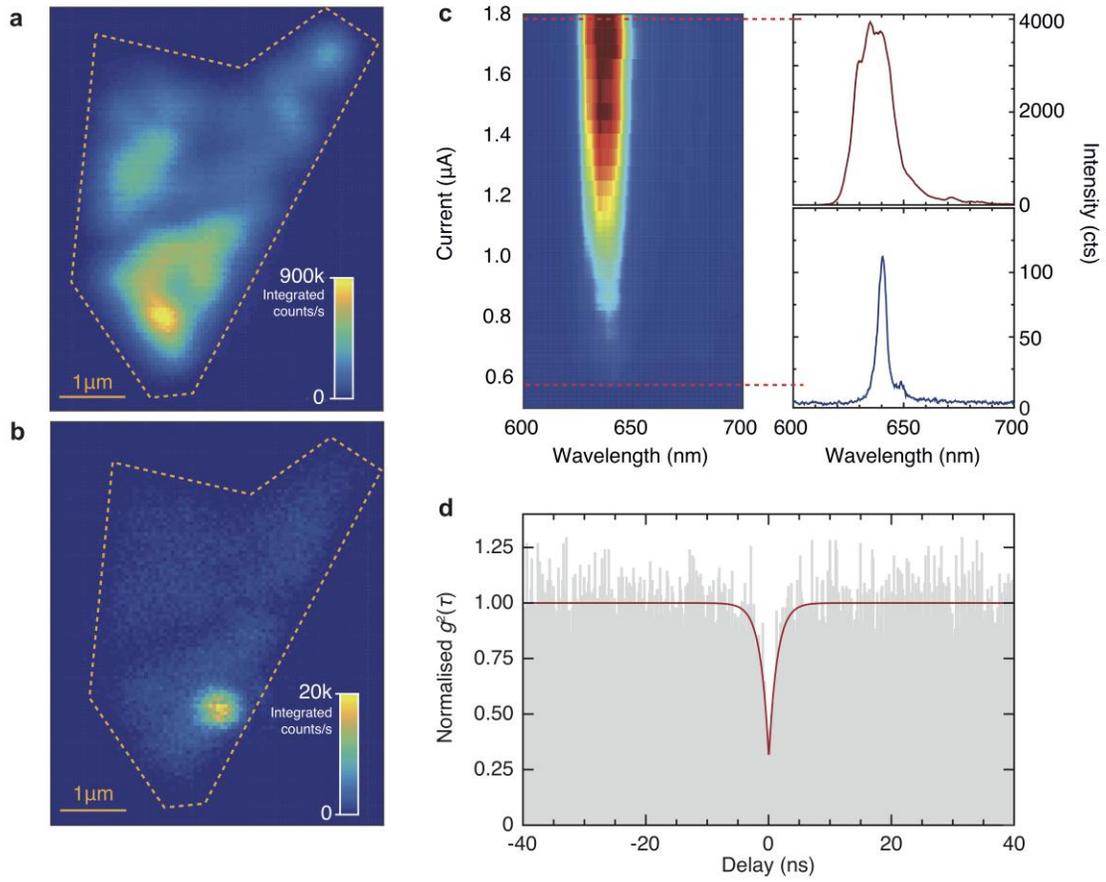